\documentclass{revtex4}
\usepackage{graphicx}
\usepackage[]{epsfig}

\newcommand{\beq}{\begin{equation}}
\newcommand{\eeq}{\end{equation}}
\newcommand{\beqd}{\begin{displaymath}}
\newcommand{\eeqd}{\end{displaymath}}
\newcommand{\beqa}{\begin{eqnarray}}
\newcommand{\eeqa}{\end{eqnarray}}

\renewcommand{\a}{\alpha}
\renewcommand{\b}{\beta}

\newcommand{\s}{\sigma}
\newcommand{\arctanh}{{\rm arctanh}}
\newcommand{\comment}[1]{}

\newcommand{\Tr}{{\rm Tr}\,}

\begin{document}

\title{Chaos in Temperature in Diluted Mean-Field Spin-Glass}

\author{Giorgio Parisi$^{1,2}$ and Tommaso Rizzo}

\affiliation{$^{1}$Dipartimento di Fisica, Universit\`a di Roma ``La Sapienza'', 
P.le Aldo Moro 2, 00185 Roma,  Italy
\\
$^{2}$Statistical Mechanics and Complexity Center (SMC) - INFM - CNR, Italy
}

\begin{abstract}
We consider the problem of temperature chaos in mean-field spin-glass models defined on random lattices with finite connectivity.
By means of an expansion in the order parameter we show that these models display a much stronger chaos effect than the fully connected Sherrington-Kirkpatrick model with the exception of the Bethe lattice with bimodal distribution of the couplings.
\end{abstract}

\maketitle

\section{Introduction}

Chaos in temperature is very old problem in spin-glass theory and has received a lot of attention over the years in connection with many different problems.
It has been studied by various approaches including: i) scaling arguments
and  real space renormalization group analysis \cite{C1}, analytical and numerical studies
on mean-field models \cite{BC,BM,Rchaos2,RY,Rchaos1} (see also \cite{Rizzolesson} for a recent review), analytical and numerical studies on the finite-dimensional Edwards-Anderson model \cite{C3,Kondor,KV},
analytical and numerical studies on elastic manifolds in random media \cite{C4}.
The problem of the temperature dependence of the Gibbs measure of the Random Energy Model has been also investigated both in the physics literature \cite{SB} and in the mathematical physics literature \cite{Kurkova}. 
It is also believed that the chaos picture is suitable to understand the surprising rejuvenation and memory effects observed in the dynamics of real spin-glasses, see for example \cite{C41,C42,C5,C6,C65,C7} and references therein.
Furthermore, a detailed understanding of this problem would probably shed light on the success of the Parallel Tempering procedure, which is nowadays considered an essential ingredient to achieve thermalization in numerical spin-glass simulations \cite{PT,Janus}. 

In this paper we consider the problem of chaos in temperature in various mean-field spin-glass models with finite connectivity. We show that these models display a much more pronounced chaos effect with respect to the fully connected Sherrington-Kirkpatrick (SK) model, with the notable exception of Bethe lattices with bimodal distribution of the couplings; moreover we will analyze the dependence of chaos on the various parameters.  Chaos is much more pronounced if the system is locally heterogenous. It is possible that some of these results also hold in finite dimensional models.

\section{Some definitions}

In the problem of temperature chaos one is interested in the correlations between the thermodynamically relevant configurations at different temperatures for a given a general spin-glass model defined by a random Hamiltonian of $N$ spins $H_J\{\sigma\}$.
In particular one would like to evaluate the probability $P_J^{\beta_1 \beta_2}(q)$ of observing an overlap $q$ if we extract two configurations according to their Boltzmann weights from systems with the same quenched Hamiltonian but different temperatures:
\beq
P_J^{\beta_1 \beta_2}(q)= \frac{\sum_{\{\tau\}\{\sigma\}}\delta( N\,q-\sum_i \sigma_i \tau_i)\exp[-\beta_1 H_J\{\sigma\} -\beta_2 H_J\{\tau\}]}{\sum_{\{\tau\}\{\sigma\}}\exp[-\beta_1 H_J\{\sigma\} -\beta_2 H_J\{\tau\}]}
\eeq  
where $\beta_i\equiv T_i^{-1}$ and the overline represents average with respect to different Hamiltonians $H_J\{\tau\}$.

A peculiar feature of Replica-Symmetry-Breaking theory \cite{MPV} is that if the two systems have the same temperature the function $P_J^{T_1T_1}(q)$ in the low temperature phase has a support  between $-q_{EA}$ and $q_{EA}$, where $q_{EA}$ is the so-called Edwards-Anderson parameter (for a two spin interaction in absence of magnetic field). In particular the disorder average $P(q)=\overline{P_J(q)}$ is given by $P(q)=dx/dq$ where $q(x)$ is a continuous function between zero and $q_{EA}$. 

The problem of chaos in temperature concerns  the function $P_J^{\beta_1 \beta_2}(q)$, in particular we say that there is chaos if 
\beq
P_J^{\beta _1 \beta_2}(q)=\delta(q)
\eeq
{\it i.e.} if $P_J^{\beta_1 \beta_2}(q)$ has a support concentrated on $q=0$ and that there is no chaos otherwise. In the nutshell, if chaos is present, the equilibrium configurations at one temperature are quite different from those at a different temperature. It is clear that chaos may have a dramatic effect of the dynamics after a temperature shift.

The problem of chaos is intrinsically related to the disordered nature of these systems, being trivial in non-disordered models like a ferromagnet that have a translational invariant order parameter. 
The problem of chaos in temperature in the various mean-field models has been investigated intensively over the years.
Today we know that there are full-RSB models that do not display chaos in temperature \cite{Rchaos1} and full-RSB models that do have chaos including notably the SK model \cite{Rchaos2}.
Similarly there are 1RSB models that do have chaos in temperature and models that do not \cite{RY}. 

If chaos is present, we would like to quantify it, also to understand the finite volume effects (or finite time effects in the dynamics). More precisely  we would like to know the free energy increase that happens if we constrain one system near to the other (or more generally at an overlap $q$). This free energy increase can (as usually) written also as large deviations function for the distribution of the overlap between systems at different temperatures:  it can be computed  by studying two coupled systems.

The free energy of two systems forced to stay at a fixed overlap $q$ is given by:
\begin{displaymath}
F_{12}(q,\beta_1,\beta_2)=-{1 \over N} \overline{ \ln \sum_{\{\tau\}\{\sigma\}}\delta\left(N\,q-\sum_i \sigma_i \tau_i\right)\exp[-\beta_1 H_J\{\sigma\} -\beta_2 H_J\{\tau\}]} \ .
\end{displaymath}
The function $F_{12}(q,\beta_1,\beta_2)$ must be larger than or equal to the free energies of the two unconstrained systems and the relevant quantity is the free energy shift $\Delta F_{12}(q,\beta_1,\beta_2) = F_{12}(q,\beta_1,\beta_2)-F(\beta_1)-F(\beta_2)$. Indeed if this quantity is greater than zero it follows that the large deviations of the overlap are given by:  
\beq
P^{\beta_1 \beta_2}_J(q) \propto \exp[- N \Delta F_{12}(q,\beta_1,\beta_2)] \ .
\eeq
If the function $P^{\beta_1 \beta_2}_J(q)$ has support on some non-zero values of $q$ then the free energy shift must vanish $\Delta F_{12}(q,\beta_1,\beta_2) = 0$. The opposite in general is not true, {\it i.e.} a vanishing free energy difference does not necessarily imply a non-zero $P^{\beta_1 \beta_2}_J(q)$ as was unexpectedly discovered  in the case $\beta_1=\beta_2$ for the spherical SK model \cite{TALA2,Rizzolesson}.

In the following we will consider the constrained free energy functional $F_{12}(q,\beta_1,\beta_2)$ averaged over the disorder because it is usually assumed that this quantity, (and correspondingly the large deviations) does not fluctuate in the large $N$ limit.  We could also define $F^Q_{12}(q,\beta_1,\beta_2)$ the free energy of the replica $\sigma$ if we constrain the replica $\sigma$ to stay at a fixed overlap from replica $\tau$, the replica $\tau$ being at equilibrium \cite{RECIPES}:
\begin{displaymath}
F^Q_{12}(q,\beta_1,\beta_2)=-{1 \over N} \overline{ \sum_{\{\tau\}}\exp[-\beta_2 H_J\{\tau\}]\ln \sum_{\{\sigma\}}\delta\left(N\,q-\sum_i \sigma_i \tau_i\right)\exp[-\beta_1 H_J\{\sigma\}] 
\over\sum_{\{\tau\}}\exp[-\beta_2 H_J\{\tau\}] } \ .
\end{displaymath}

In the first definition everything was symmetric in the two replicas and forcing the system to have a non zero overlap we push out of equilibrium {\em both} replicas.
On the contrary in the second definition we look to the probability of $\sigma$, when it is constrained to stay at a fixed overlap with $\tau$ is a quenched configuration at equilibrium.
It is evident that convexity implies that
\begin{displaymath}
F_{12}(q,\beta_1,\beta_2)<F^Q_{12}(q,\beta_1,\beta_2)\ ,
\end{displaymath}
so that if $F_{12}(q,\beta_1,\beta_2)$ displays chaos, chaos is present also in $F^Q_{12}(q,\beta_1,\beta_2)$. It is interesting that we can obtain the internal energy as function of $q$ by performing a derivative with respect to $\beta_2$. In presence of chaos the quantity
\beq
\Delta E(\beta_1,\beta_2)=E^Q_{12}(q_{EA},\beta_1,\beta_2)-F^Q_{12}(0,\beta_1,\beta_2) \ ,
\eeq
should have the meaning of the energy that is {\em slowly} released after a sudden quench from a high temperature. Unfortunately these slow relaxations are too small to be observed experimentally.

In this paper we compute $F_{12}(q,\beta_1,\beta_2)$, the computation of $F^Q_{12}(q,\beta_1,\beta_2)$ could be cone in a similar way: the two functions display a similar qualitative behaviour.

In order to compute the free energy shift $\Delta F_{12}(q,\beta_1,\beta_2)$ is is convenient to consider the following {\it coupled} free energy:
\beq
\tilde{F}_{12}(\epsilon, \beta_1,\beta_2 )=-{1 \over N}\overline{\ln \sum_{\{\tau\}\{\sigma\}}\exp\left[-\beta_1 H_J\{\sigma\}-\beta_2 H_J\{\tau\}+\epsilon \sum_{i=1}^N \sigma_i \tau_i \right]}
\eeq
where  $\beta_i$ is the inverse temperature of the corresponding system and $N$ is the total number of spins in the system.
When the coupling term $\epsilon$ vanishes the above quantity is simply the sum of the averages free-energies (times $\beta$) at inverse temperatures $\beta_1$ and $\beta_2$, therefore the relevant physical information concerning chaos is given by the difference $\Delta \tilde{F}_{12}(\epsilon,\beta_1,\beta_2)=\tilde{F}_{12}(\epsilon, \beta_1,\beta_2 )-F_1(\beta_1)-F_2(\beta_2)$. 
In the thermodynamic limit the two functions $F_{12}(q,\beta_1,\beta_2)$ and  $\tilde{F}_{12}(\epsilon,\beta_1,\beta_2)$
 becomes Legendre transform of each other through the following relationships:
\beq
\tilde{F}_{12}(\epsilon) = F_{12}(q^*)- q \epsilon\, ,  \ \ \ \ \ \ \ \ \ \ \ \  
\left.{d F \over dq}\right|_{q=q^*}= \epsilon
\eeq
and 
\beq
F_{12}(q) =\tilde{F}_{12}(\epsilon^*) +q \epsilon\, ,  \ \ \ \ \ \ \ \ \ \ \ \  
\left.-{d \tilde{F} \over d\epsilon}\right|_{\epsilon = \epsilon^*}= q.
\eeq

\section{Chaos in Temperature in the Generalized SK model}
\label{chaosgeneral}

In this section we consider the problem of chaos in temperature in the context of the generalized SK model, that is a spin-glass model whose free energy is given by the extremization of a functional of an $n \times n$ matrix $Q_{ab}$ with generic coefficients. Over the years many different spin-glass models have been mapped over a generalized Sherrington-Kirkpatrick model {\it e.g.} the Edwards-Anderson model in an expansion in large dimension \cite{GMY} and at fixed dimension in the loop expansion above the upper critical dimension $D_u=6$ \cite{KDT}.
In particular such a mapping was used by Kondor to assess chaos in temperature in the Edwards-Anderson model \cite{Kondor,KV}. In the next section we will use it to study spin-glass models on the Bethe lattice.

In all the aforementioned spin-glass models $F_{12}(\epsilon, \beta_1,\beta_2 )$ can be computed in the replica framework (see {\it e.g.} \cite{Rizzolesson}), {\it i.e.} considering a set of $n$ replicas of the two coupled systems and then sending $n$ to zero.
By means of standard manipulations one obtains the free energy as a functional $F_{12}(\hat{Q},\epsilon)$ over a $ 2n \times\ 2n$ matrix
 $\hat{Q}=\left(\begin{array}{cc}
 Q_{1} & P \\
 P^{t} & Q_{2}
 \end{array}\right)$, where $Q_1,Q_2$ and $P$ are $n \times n$ matrices.
The constrained free energy is obtained extremizing the functional with respect to the order parameter $\hat{Q}$ at a given value of $\epsilon$.
The functional may be very complicated but one can get a more tractable expression expanding it in powers of $\hat{Q}$. This expansion is perturbative near the critical temperature where $\hat{Q}$ is expected to be small. In the end one obtains the following variational expression:
\beqa
 F_{12}(\hat{Q},\epsilon) & = & F_{\rm para}(\beta_1)+F_{\rm para}(\beta_2)+
\nonumber
\\
& - &{\tau_1\over 2}\Tr Q_1^2-{\tau_2\over 2}\Tr Q_2^2-{\tau_{12}}\Tr P^2-{\omega \over 6} \Tr \hat{Q}^3 +
\nonumber
\\
& - & {v \over 8} \Tr \hat{Q}^4 + {y \over 4} \sum_{abc}\hat{Q}^2_{ab}\hat{Q}^2_{ac}-{u \over 12}\sum_{ab}\hat{Q}_{ab}^4 - \epsilon \,  \sum_{a=1}^n P_{aa}+O(\hat{Q}^5),
\label{fvar}
\eeqa
where  $F_{\rm para}(\beta_1)$ and $F_{\rm para}(\beta_2)$ are terms that do not depend on $\hat{Q}$  and are irrelevant for the present discussion.
The above expression is $O(n)$ and we have neglected terms $O(n^2)$ that can be present and are relevant to evaluate the free energy fluctuations \cite{PRdil}. 
For the time being we assume that the only dependence on the temperature is in the reduced temperature $\tau_1$, $\tau_2$ and $\tau_{12}$ while the other coefficients do not change with the temperature. 

The variational action in presence of a forcing term $\epsilon$ has been computed in \cite{mio1} and will be also reconsidered in appendix \ref{quadratic}. It turns out that the chaos effect depends crucially on the parameters $\omega$, $v$ and $c_{12}$ that is the coefficient of the term $(\tau_1-\tau_2)^2$ in the expansion of $\tau_{12}$:
\beq
\tau_{12}={1 \over 2}(\tau_1+\tau_2)+ {c_{12} \over 4}(\tau_1-\tau_2)^2+O(\tau^3).
\label{eqt12}
\eeq
Following \cite{mio1,Rizzolesson} we report the following value of the free energy shift at leading order in $q$ and $(\tau_1-\tau_2)$:
\beq
\Delta F_{12}(q)= A\, {|q|^3}(\tau_1-\tau_2)^2 \ \ \ A={ u \over 6 \omega }\left({v \over \omega^2}-c_{12}\right) \ , 
\label{DFq3}
\eeq 
The above expression holds when $q$ is small but it is larger than $|\tau_1-\tau_2|$; in the opposite situation ($q<<\tau_1-\tau_2|$), as we will show in appendix \ref{quadratic}, we have at leading order:
\beq
\Delta F_{12}(q)=B\, q^2|\tau_1-\tau_2|^3\ \ \ 
B={u^{1/2} \over   \omega 2^{3/2} \pi}\left({v \over \omega^2}-c_{12}\right)^{3/2}
\label{DFq2}
\eeq
A peculiar feature of the SK model is that the quantity $\left({v \over \omega^2}-c_{12}\right)$ vanishes because $\omega=v=c_{12}=1$ therefore chaos is not present at this order. In this case a more refined computation is necessary  \cite{Rchaos2} and it shows that relationship (\ref{DFq3}) must be replaced with the much smaller expression
\begin{equation}
\Delta F_{12}(q) ={12 \over 35}\ |q|^7\Delta T^2 \,.
\end{equation}
As a consequence chaos in temperature in the SK model is exceedingly weak and it was not observed in numerical simulations up to quite large system sizes \cite{BM}.

\section{Chaos in Temperature on Bethe Lattice Spin-Glass Models}
\label{ChaosBethe}

\subsection{Setting up the computation} 
In \cite{PRdil} we have obtained the mapping of the free energy of spin-glass models defined on Bethe lattices with finite connectivity on the action (\ref{fvar}). In this section we will extend those results to study chaos in temperature in these models.

Extending the treatment of \cite{GDD} to the case of two coupled systems we express the free energy as a  variational functional of the order parameter $\rho(\{\sigma_1,\sigma_2\})$ that is a function defined on $2n$ Ising spins $\{ \s_1\} \equiv \sigma_1^1, \dots,\sigma_1^n$ and $\{ \s_2\} \equiv \sigma_2^1, \dots,\sigma_2^n$.
The variational expression of the free energy reads:
\beqa
\tilde{F}_{12}(\epsilon,\beta_1,\beta_2) & = & {M \over n} \ln \Tr_{ \{ \s_1, \s_2 \} } \rho^{M+1}\{ \s_1,\s_2  \}e^{\epsilon \sum_{a=1}^n \s_1^a\s_2^a}+
\nonumber
\\
& - & \frac{M+1}{2n}\ln \int  \Tr_{ \{ \s_1 ,\s_2 \}} \Tr_{ \{ \tau_1, \tau_2 \}} \rho^{M}\{ \s_1,\s_2  \} \rho^{M}\{ \tau_1,\tau_2  \} \times
\nonumber
\\
& \times & \left\langle  \exp \left[ \beta_1 J \sum_{\a}\s^{\a}_1\tau^{\a}_1+\beta_2 J \sum_{\a}\s^{\a}_2\tau^{\a}_2 +
\epsilon \sum_{a=1}^n \s_1^a\s_2^a + \epsilon \sum_{a=1}^n \tau_1^a\tau_2^a \right]
\right\rangle
\label{PHIdil}
\eeqa
Where $M+1$ is the connectivity of the lattice and the square brackets mean average with respect to the distribution of $J$.
The above expression has to be extremized with respect to $\rho\{ \s_1,\s_2  \}$. We note that it is invariant under a rescaling of $\rho\{ \s_1,\s_2  \}$ so that we can choose any normalization for it.
If we normalize $\rho\{ \s_1,\s_2  \}$ to one the corresponding variational equation in terms of $\rho(\sigma)$ reads:
\beq
\rho\{ \s_1,\s_2  \}= {1 \over {\mathcal N}}\Tr_{ \{ \tau_1, \tau_2 \}} \rho^{M}\{ \tau_1,\tau_2  \} 
\left\langle  \exp \left[ \beta_1 J \sum_{\a}\s^{\a}_1\tau^{\a}_1+\beta_2 J \sum_{\a}\s^{\a}_2\tau^{\a}_2 + \epsilon \sum_{a=1}^n \tau_1^a\tau_2^a \right]
\right\rangle
\label{varDIL}
\eeq
where ${\mathcal N}$ is a normalization constant.

In order to build an expansion in the order parameter we write:
\beq
\rho\{ \s_1,\s_2  \} = \sum_{k_1=0,k_2=0}^n b_{k_1,k_2} \sum_{(\a_1 \dots \a_{k_1}),(\b_1 \dots \b_{k_2})}q_{\a_1\dots\a_{k_1},\b_1 \dots \b_{k_2}}\s_1^{\a_1}\dots \s_1^{\a_{k_1}}\s_2^{\b_1}\dots \s_2^{\b_{k_2}}
\eeq
with 
\beq
b_{k_1,k_2} \equiv \langle \cosh^n \beta_1 J \cosh^n \beta_2 J \tanh^{k_1} \beta_1 J \, \tanh^{k_2} \beta_2 J\rangle 
\label{defb}
\eeq
The variational equation (\ref{varDIL}) can now be written as equations for $q_{\a_1\dots\a_{k_1},\b_1 \dots \b_{k_2}}$:
\beq
q_{\a_1\dots\a_{k_1},\b_1 \dots \b_{k_2}} = { \Tr_{ \{ \tau_1, \tau_2 \}}  \tau_1^{\a_1}\dots \tau_1^{\a_{k_1}} \tau_2^{\b_1}\dots \tau_2^{\b_{k_2}} \rho^{M}\{ \tau_1,\tau_2  \} \exp \epsilon \sum_{a=1}^n \tau_1^a\tau_2^a 
\over
\Tr_{ \{ \tau_1, \tau_2 \}}  \rho^{M}\{ \tau_1,\tau_2  \} \exp \epsilon \sum_{a=1}^n \tau_1^a\tau_2^a }
\label{equaq}
\eeq
The r.h.s. of the above equations can be expanded in powers of $q_{\a_1\dots\a_{k_1},\b_1 \dots \b_{k_2}}$. 

In general the equations for two-index $q_{ab}$ depend on higher order objects and this fact leads to rather complex equations. However near the critical temperature the $Q$ with an high number of indices terms are much smaller that the one with two indices and using the corresponding variational equations they can be eliminated \cite{PARISITRIA}. In this way  we can obtain an expression that depends only on a $2n \times 2n$ matrix  $\hat{Q}$ like the one appearing in the variational action (\ref{fvar}). This was done in \cite{PRdil} for the single system and we have extended that computation to the coupled system case.

If we start from the equations in appendix B of \cite{PRdil}, we divide  all the equations by a factor 2 and 
 we make a rescaling $Q_1 \rightarrow Q_1 /(b_{11} (M-1))$, $Q_2 \rightarrow Q_2 /(b_{22}(M-1))$, $P \rightarrow P /(b_{12}(M-1))$,  we obtain that the equations (\ref{equaq}) for  $\hat{Q}$ are the same that would be obtained from a variational action of the form (\ref{fvar}) with the following coefficients:  
\beqa
\tilde{\tau}_1 & = & {M b_{11}-1 \over 2 b_{11} (M-1)}
\\
\tilde{\tau}_2 & = & {M b_{22}-1 \over 2 b_{22} (M-1)}
\\
\tilde{\tau}_{12} & = & {M b_{12}-1 \over 2 b_{12} (M-1)}
\\
\tilde{\omega} & = &  {M \over M-1}
\\
\tilde{v} & = & {M(M b_4+M-2) \over (1-M b_4)(M-1)^2} 
\\
\tilde{u} & = &{M(M(2M-1)b_4+M-2)\over  (1-M b_4)(M-1)^2}
\eeqa
where $b_4=\langle (\tanh \beta_c J)^4\rangle$ and $\beta_c$ is the inverse critical temperature that obeys the equation $1=M \langle \tanh^2 \beta_c J\rangle$.
The mixed reduced temperature $\tilde{\tau}_{12}$ is such that its expansion in terms of $\tilde{\tau}_1$ and $\tilde{\tau}_2$ is of the form (\ref{eqt12}) with the following expression for $c_{12}$:
\beq
\tilde{c}_{12}={2(M-1) \over M}-{ (M-1) \overline{J^2 (1-\tanh^2 \beta_c J)^2}\over  (\overline{J \tanh\beta_c J}-\overline{J \tanh^3\beta_c J})^2 M^2}
\eeq
The Sherrington-Kirkpatrick limit is recovered sending $M$ to infinity and the coupling strength to zero as $\overline{J^2}=1/M$. In this limit we have $b_{ij}=1/(M T_i T_j)$, the critical temperature goes to 1 and the various coefficients read:
\beq
\tilde{\tau}_1  =  {1-T_1^2 \over 2 }\,,\ \ \tilde{\tau}_2  =  {1-T_2^2 \over 2 }\,,\ \ \tilde{\tau}_{12}  =  {1-T_1 T_2 \over 2 }\,,
\eeq
\beq
\tilde{u}=\tilde{\omega}=\tilde{v}=\tilde{c}_{12}=1.
\eeq
These are precisely the coefficients obtained for the SK model, see {\it e.g.} \cite{mio1}.
The rescaling $Q_1 \rightarrow Q_1 /(b_{11} (M-1))$, $Q_2 \rightarrow Q_2 /(b_{22}(M-1))$, $P \rightarrow P /(b_{12}(M-1))$ was performed in order to get rid of the temperature dependence in all coefficients other than $\tilde{\tau}_1$, $\tilde{\tau}_2$ and $\tilde{\tau}_{12}$ and corresponds in the SK limit to the usual rescaling $Q_1 \rightarrow Q_1 /\beta_1^2$, $Q_2 \rightarrow Q_2 /\beta_2^2$, $P \rightarrow P /(\beta_1 \beta_2)$.
The above definitions are such that at finite $M$ the reduced temperature goes to $1/2$ at zero temperature. The actual dependence of the reduced temperature with respect to the temperature is such that:
\beq
\tilde{\tau}={(\overline{J \tanh\beta_c J}-\overline{J \tanh^3\beta_c J})M^2 \over (M-1) T_c^2}(T_c-T)+O(T_c-T)^2
\label{tautemp}
\eeq 
and the prefactor goes to 1 in the SK limit.

The above coefficients however cannot be put simply into eqs. (\ref{DFq3}) and (\ref{DFq2}) in order to obtain the free energy shifts. We must bear in mind that once the variational equations (\ref{equaq}) are expanded in powers of $\hat{Q}$ they {\it look like} as if they where obtained from a variational free energy of the form (\ref{fvar}) with coefficients that in order to avoid possible confusion we represent as tilded.
This does not means that the true free energy has an expansion with the same coefficients; as it was shown in appendix B of \cite{PRdil}  this can be understood noticing that the equation for the order parameter corresponds to the following expression:
\beq
0=\Tr \left[ \s_a \s_b \left( \rho(\s) - \frac{\Tr_{\tau} \rho^M(\tau)\langle \exp J \sum_c \s_c \tau_c \rangle }{\Tr \rho^M(\tau)} \right) \right]
\label{op}
\eeq
while the equation one obtains by differentiating expression (\ref{PHIdil}) corresponds to:
\beq
0=\Tr \left[ \rho^{M-1}(\{\s\}) \s_a \s_b \left( \rho(\s) - \frac{\Tr_{\tau} \rho^M(\tau)\langle \exp J \sum_c \s_c \tau_c \rangle }{\Tr \rho^M(\tau)} \right) \right] \ .
\eeq
Thus the two expressions are equivalent in the sense that they have the same solution at the order at which they are valid. In order to obtain the expansions in powers of the order parameter matrix one could expand directly the free energy (\ref{PHIdil}), but technically it is much simpler to expand the variational equations (\ref{equaq}).

This problem can be bypassed noticing that the derivatives of expressions (\ref{DFq3}) and (\ref{DFq2}) {\it with the tilded coefficients} allows to determine $q$ as a function of $\epsilon$. In the two different regimes we have:
\begin{equation}
{3 \, q^2 \tilde{u} \over 6 \, \tilde{\omega} }\left({ \tilde{v} \over \tilde{\omega}^2}-\tilde{c}_{12} \right) ( \tilde{\tau}_1-\tilde{\tau}_2 )^2=\epsilon
\label{eqe1}
\end{equation}
and
\beq
2 q{\tilde{u}^{1/2} \over   \tilde{\omega} 2^{3/2} \pi}\left({\tilde{v} \over \tilde{\omega}^2}-\tilde{c}_{12}\right)^{3/2}|\tilde{\tau}_1-\tilde{\tau}_2|^3= \epsilon
\label{eqe2}
\eeq
We must take into account that the overlap $q$ appearing in the above equation is not the true overlap. This is due to two reasons:
\begin{itemize}
\item
We must recall that we have done the rescaling $Q_1 \rightarrow Q_1 /(b_{11} (M-1))$, $Q_2 \rightarrow Q_2 /(b_{22}(M-1))$, $P \rightarrow P /(b_{12}(M-1))$ 
\item
A more subtle reason is that  the true overlap is given by:
\beq
q = { \Tr_{ \{ \tau_1, \tau_2 \}}  \tau_1^{1}\tau_2^{1}\rho^{M+1}\{ \tau_1,\tau_2  \} \exp \epsilon \sum_{a=1}^n \tau_1^a\tau_2^a 
\over
\Tr_{ \{ \tau_1, \tau_2 \}}  \rho^{M+1}\{ \tau_1,\tau_2  \} \exp \epsilon \sum_{a=1}^n \tau_1^a\tau_2^a }\ .
\eeq
The difference is that there is a term  $\rho^{M+1}$ while in equation (\ref{equaq}) there is a power  $\rho^{M}$. As usual the overlap entering in the cavity equations is not the true overlap.
\end{itemize}
The net effect is that in order to obtain {\it at leading order} the relationship between the true overlap and the forcing $\epsilon$ we have to make the following rescaling in eqs. (\ref{eqe1}) and (\ref{eqe2}), 
\beq
q \rightarrow {b_{12}(M-1) \over 1+b_{12}}q \ .
\eeq
In the SK limit the above rescaling reduce to $q \rightarrow \beta_1 \beta_2 q$. Near the critical temperature we have:
\beq
{b_{12}(M-1) \over 1+b_{12}}={M-1 \over M+1}
\eeq 
The corresponding expressions yield the overlap as a function of the forcing and can be integrated back to get the correct free energy shifts in the two regimes $(\tilde{\tau}_1-\tilde{\tau}_2) \ll q$ and $q \ll (\tilde{\tau}_1-\tilde{\tau}_2)$:
\begin{equation}
\Delta F_{12}(q)= A\, |q|^3 (\tilde{\tau}_1-\tilde{\tau}_2)^2 \ \ \ 
A=\left({M-1 \over M+1}\right)^2{ \tilde{u} \over 6 \tilde{\omega} }\left({\tilde{v} \over \tilde{\omega}^2}-\tilde{c}_{12}\right) \, ,
\label{DFq3tilde}
\end{equation}
and
\beq
\Delta F_{12}(q)=B\, q^2|\tilde{\tau}_1-\tilde{\tau}_2|^3\ \ \  \ B=\left({M-1 \over M+1}\right){\tilde{u}^{1/2} \over   \tilde{\omega} 2^{3/2}  \pi}\left({\tilde{v} \over \tilde{\omega}^2}-\tilde{c}_{12}\right)^{3/2}
\label{DFq2tilde}
\eeq

We can now apply the previous formulae to different distributions of the $J$.

\subsection{Diluted bimodal Distribution}

A surprising feature of the above expressions is that a direct computation shows that the quantity $\tilde{v}/\tilde{\omega}^2-\tilde{c}_{12}$ vanishes in the case of a bimodal distribution of the coupling $J=\pm 1$ as in the SK model. Therefore for these models we expect chaos to be a much smaller effect possibly of the same order of the SK model. This is consistent with the fact that chaos is very difficult to be observed in numerical simulations of these models. 

This can be seen considering the case of the random-bond bimodal distribution where any coupling in the lattice is zero with probability $p$ or $\pm 1$ with probability $1-p$: 
\beq
P(J)=p\, \delta(J)+{(1-p) \over 2}(\delta(J+1)+\delta(J-1)) \ \ \ (0 \leq p \leq 1) \ .
\eeq
In this case the relevant parameters to be insert in eqs. (\ref{DFq3tilde}) and (\ref{DFq2tilde}) read:
\beq
\beta_c = \arctanh \, {1 \over \sqrt{M(1-p)}}
\eeq
\beq
\tilde{\omega} ={M \over M-1}\,, \ \ \tilde{u}={M (1-M^2(1-p)-2 M p) \over (M-1)^2 (1-M (1-p))} 
\eeq
\beq
|\tilde{\tau}_1-\tilde{\tau}_2|={ M (M(1-p)-1)\arctanh^2 {1 \over \sqrt{M(1-p)}}\over (M-1)\sqrt{M(1-p)} } |T_1-T_2|+O(T_1-T_2)^2
\eeq
In the last expression we have used eq. (\ref{tautemp}). The chaos prefactor is:
\beq
{\tilde{v} \over \tilde{\omega}^2}-\tilde{c}_{12}={p \over M(1-p)-1}
\label{1mvc}
\eeq
and we see that it vanishes in the purely bimodal case corresponding to $p=0$.

\subsection{Poissonian Distribution of the Connectivity}

The parameters computed above can be also used to study the model where the connectivity of each spin has a Poissonian distribution.  In general we have to take the $M \rightarrow \infty$ limit in the above expressions sending $p \rightarrow 1$ as 
\beq
p=1-{\alpha \over M}.
\eeq
where $\alpha$ is the average connectivity of a site.
In the case where the couplings strength is $\pm 1$ the relevant parameters read:
\beq
\beta_c=\arctanh \, {1 \over \sqrt{\alpha}}
\eeq
\beqa
\tilde{\omega} & = & 1
\\
\tilde{u} & = & {2 +\alpha \over \alpha - 1}
\\
{\tilde{v} \over \tilde{\omega}^2}-\tilde{c}_{12} & = & {1 \over \alpha - 1}
\eeqa
\beq
|\tilde{\tau}_1-\tilde{\tau}_2|={\alpha -1 \over \sqrt{\alpha}} \arctanh^2 \,{1 \over \sqrt{\alpha}}\, |T_1-T_2|+O(T_1-T_2)^2\ . 
\eeq
\beq
{\tilde{v} \over \tilde{\omega}^2}-\tilde{c}_{12}={1 \over \alpha-1}
\eeq
The free energy differences read:
\beq
\Delta F_{12}(q)=|q|^3 \,{ 2 +\alpha \over 6 \alpha} \arctanh^2 \,{1 \over \sqrt{\alpha}}\, \left({T_1-T_2\over T_c}\right)^2
\eeq 
\beq
\Delta F_{12}(q)=q^2\,{(2 +\alpha)^{1/2}(\alpha - 1) \over   (2 \alpha)^{3/2}  \pi} \arctanh^4 \,{1 \over \sqrt{\alpha}} \left|{T_1-T_2 \over T_c}\right|^3
\eeq

\section{Conclusions}

We have shown that mean-field spin-glass models defined on random lattices with finite connectivity display chaos in temperature. Chaos is stronger than in the SK model by four orders of magnitude in perturbation theory.

In general if we consider the region of small overlaps we could expect a chaos effect such as  $\Delta F_{12}(q) \propto q^2 |T_1-T_2|^2$ instead as we have seen in section (\ref{chaosgeneral}) the effect is smaller $\Delta F_{12}(q) \propto q^2 |T_1-T_2|^3$. In other words chaos in the generalized SK model, to which the random lattice models can be mapped, is larger than in SK but is nevertheless not as strong as one could naively expect. This is because in Bethe lattice models chaos is truly a RSB effect \cite{RY}: a model with a stable RS phase has a single stable thermodynamic state that can be followed increasing or decreasing the temperature and therefore is not chaotic \footnote{The Spherical SK model is RS and is indeed non-chaotic but has a zero-free energy shift at order $N$ \cite{TALA2,Rizzolesson}. However this is a consequence of the fact that the model is marginally stable in the whole low temperature phase. A different scenario is present in the Migdal-Kadanoff approximation \cite{C1}, however the presence of a few spins with very large number of connections may have deep effects on the properties of this model. }. The connection between chaos and RSB is reflected by the fact that the coefficients of the free energy shifts eqs. (\ref{DFq3}) and (\ref{DFq2}) depend on the coefficient $u$ of the quartic interaction that is responsible for RSB.

In diluted models chaos in temperature is considerably stronger than in SK with the notable exception of Bethe lattices with bimodal interactions as shown by eq. (\ref{1mvc}). In this case we expect the effect to be of the same order of magnitude of the SK model. In the SK model one can be prove that the quadratic terms in $q^2$ in the free energy shift vanishes at all orders \cite{mio1,Kondor}, we mention that the same result can be proven at all orders in the case of the Bethe lattice with purely bimodal interaction but this will be published elsewhere. The argument relies on the local homogeneity of the Bethe lattice with bimodal interactions that is also responsible of non-Gaussian free energy fluctuations \cite{BKM,PRdil}. 

The fact that chaos in temperature on Bethe lattice with bimodal interaction is as weak as in the SK model is supported also by existing numerical results. The numerical data of Billoire and Marinari (BM) suggest absence of chaos in SK \cite{BM}, only a close look at the function $P^{\beta_1\, \beta_2}(q)$ gives a hint that the effect may be present due to a very slow increasing of the weight in $P^{\beta_1\, \beta_2}(0)$ with the system size. On the other hand the theoretical value computed in \cite{Rchaos2} shows that the effect is exceedingly small in the SK model and that it is practically unobservable at the system sizes simulated in \cite{BM}. 
BM considered also the Bethe lattice with connectivity $c=6$ and bimodal distribution of the coupling finding again no strong chaos effect in agreement with the results presented here. 

Finally we would like to remark that the extension of our computations to finite dimensional models (at least for  large dimensions) can be done using the techniques of \cite{GMY}. It would be very interesting to study if we can obtain reliable predictions in high dimensional models, e.g. $D=6$.

\appendix
\section{The Quadratic Action}
\label{quadratic}

In this subsection we compute the free energy shift when $\epsilon$ is much smaller than any other parameter in the theory. In this case the problem can be solved expanding the variational action (\ref{fvar}) at second order around the solution $P_{ab}=0$. A similar treatment was recently put forward in the case of bond chaos \cite{Rizzolesson,Aspel}. The $P$-dependent term in the action can be written as:
\beq
F_{12}(P,\epsilon)={1 \over 2} \Tr P A P - \epsilon \,  \Tr P+o(P^2)
\eeq
where the matrix $A$ is given by:
\beq
A_{ab}=-(2 \tau_{12}-{y \over n}\Tr Q_1^2-{y \over n}\Tr Q_2^2)\delta_{ab}-\omega (Q_1+Q_2)_{ab}-v\,(Q_1^2+Q_2^2+Q_1Q_2)_{ab}
\eeq
extremizing with respect to $P$ the above expression at given $\epsilon$ we easily obtain:
\beq
F_{12}(\epsilon,\beta_1,\beta_2)=-{ \epsilon^2 \over 2}\Tr A^{-1} 
\eeq
In order to compute the trace we diagonalize the matrix $A_{ab}$. The eigenvalues of a hierarchical matrix described by $(a_d,a(x))$ are given by
\begin{eqnarray}
\lambda_0 & = & a_d-\int_n^1\,a(x)dx \ \ \ \ \ \ \ \ \ \ \ \ \ \ \ \ \ \ \ {\rm deg:}\ \ 1
\nonumber
\\
\lambda_x & = & a_d-\left(x a(x)+\int_x^1q(x)dx \right) \ \ \ {\rm deg:}\ -n{dx \over x^2}
\nonumber
\end{eqnarray}
and the trace is given by:
\beq
\lim_{n \rightarrow 0}{1 \over n}\sum_a {1 \over\lambda_a}={1 \over\lambda_1}-\int_0^1\dot{a}(y){1 \over \lambda(y)^2}
\eeq
Let us examine the eigenvalue $\lambda_a(0)$:
\beq
-\lambda_a(0)=2 \tau_{12}-{y \over n}\Tr Q_1^2-{y \over n}\Tr Q_2^2-\omega(\overline{q}_1+\overline{q}_2)+v(\overline{q}_1^2+\overline{q}_2^2+\overline{q}_1\overline{q}_2)
\eeq
at this point we exploit the fact that when $\tau_1=\tau_2$ this eigenvalue must vanish obtaining the condition:
\beq
\tau_{1}-{ y \over n}\Tr Q_1^2- \omega \overline{q}_1+{3 \over 2} v \overline{q}_1^2=0
\eeq
summing the above equation for $\tau_1$ and $\tau_2$ to the expression of $\lambda_a(0)$ we obtain
\beq
\lambda_a(0)=-2 \tau_{12}+\tau_1+\tau_2+{v \over 2}(\overline{q}_1-\overline{q}_2)^2=-{1 \over 2}\left(c_{12}-{v \over \omega^2}\right)(\tau_1-\tau_2)^2
\eeq
Thus we encounter the same factor $\left(c_{12}-{v \over \omega^2}\right)$ of expression (\ref{DFq3}).
In order to complete the computation of the trace we need the expression of $\lambda_a(x)$ around $x=0$. In order to do this we use the relationship $\dot{\lambda}_a(x)=-x \dot{a}(x)$, where the dot means derivative with respect to $x$.
At leading order in $\tau_1$ and $\tau_2$ we have $a(x)=-\omega (q_1(x)+q_2(x))$. The solution of the free problem is such that $q_1(x)=q_2(x)=\omega x/2 u$ \cite{MPV}, therefore 
\beq
\lambda_a(x)=\lambda_a(0)+{\omega^2 \over 2 u}x^2 \longrightarrow (\lambda_a(x))^2=\lambda_a^2(0)+\lambda_a(0){\omega^2 \over   u}x^2
\eeq
thus at leading order for small $\lambda_a(0)$ we have:
\beq
\lim_{n\rightarrow 0}{1 \over n}\Tr A^{-1}= \int_0^1 {\omega^2\over u}{dx \over \lambda_a^2(0)+\lambda_a(0){\omega^2 \over  u}x^2} \simeq {\omega \over u^{1/2}\lambda_a^{3/2}(0)}\int_0^{\infty}{du \over 1+y^2}={\omega \pi \over 2 u^{1/2}\lambda_a^{3/2}(0)}
\eeq
therefore
\beq
\Delta F_{12}(\epsilon,\beta_1,\beta_2)=-{\omega  \epsilon^2\pi \over 4 u^{1/2}\lambda_a^{3/2}(0)}
\eeq
and 
\begin{equation}
\Delta F_{12}(q,\beta_1,\beta_2)=q^2{u^{1/2} \over   \omega 2^{3/2} \pi}\left(c_{12}-{v \over \omega^2}\right)^{3/2}|\tau_1-\tau_2|^3
\end{equation}

\end{document}